# Electric-field switchable chirality in rhombohedral graphene Chern insulators stabilized by tungsten diselenide

Jing Ding[1,2,*], Hanxiao Xiang[1,2,*], Jiannan Hua[1,2,*], Wenqiang Zhou[1,2], Naitian Liu[1,2], Le Zhang[1,2], Na Xin[3], Bing Wu[4], Kenji Watanabe[5], Takashi Taniguchi[6], Zdeněk Sofer[4], Wei Zhu[1,2], Shuigang Xu[1,2,†]

[1] *Key Laboratory for Quantum Materials of Zhejiang Province, Department of Physics, School of Science, Westlake University, 18 Shilongshan Road, Hangzhou 310024, Zhejiang Province, China*

[2] *Institute of Natural Sciences, Westlake Institute for Advanced Study, 18 Shilongshan Road, Hangzhou 310024, Zhejiang Province, China*

[3] *Department of Chemistry, Zhejiang University, Hangzhou, 310058, China*

[4] *Department of Inorganic Chemistry, University of Chemistry and Technology Prague, Technická 5, 166 28, Prague 6, Czech Republic*

[5] *Research Center for Electronic and Optical Materials, National Institute for Materials Science, 1-1 Namiki, Tsukuba 305-0044, Japan*

[6] *Research Center for Materials Nanoarchitectonics, National Institute for Materials Science, 1-1 Namiki, Tsukuba 305-0044, Japan*

[*]These authors contributed equally to this work.
[†]Correspondence to: xushuigang@westlake.edu.cn

**Abstract:**
Chern insulators host topologically protected chiral edge currents with quantized conductance characterized by their Chern number. Switching the chirality of a Chern insulator, namely, the direction of the edge current, is highly challenging due to topologically forbidden backscattering but is of considerable importance for the design of topological devices. Nevertheless, this can be achieved by reversing the sign of the Chern number through a topological phase transition. Here, we report electrically switchable chirality in rhombohedral multilayer graphene-based Chern insulators. By introducing moiré superlattices in rhombohedral heptalayer graphene, we observed a cascade of topological phase transitions at quarter electron filling of a moiré band with the Chern number tunable from -1, 1 to 2. Furthermore, integrating monolayer tungsten diselenide at the moiréless interface of rhombohedral decalayer graphene/h-BN superlattices stabilizes the Chern insulators, enabling quantized anomalous Hall resistance of $h/2e^2$. Remarkably, the Chern number can be switched from -1 to 2 using displacement fields. Our work establishes rhombohedral multilayer graphene moiré superlattices as a versatile platform for topological engineering, with switchable chirality offering significant promise for integrating chiral edge currents into topological electronic circuits.

# I. INTRODUCTION

The interplay between nontrivial band topology and strong correlations produces exotic quantum states, such as quantum anomalous Hall insulators [1,2]. A quantum anomalous Hall insulator can support topologically protected dissipationless edge states, similar to those in the quantum Hall effect, but without requiring an external magnetic field. This property has promising applications in low-power-consumption electronics and topological quantum computers [3,4]. The quantum anomalous Hall insulator is characterized by a quantized Hall resistance $R_{xy} = \frac{h}{Ce^2}$ accompanied by a vanishing $R_{xx}$, where $h$ is the Planck constant, $e$ is the elementary charge, and $C \neq 0$ is the Chern number. Experimentally, the incipient quantum anomalous Hall insulator can be identified through the observation of a Chern insulator under finite magnetic fields. The Chern number $C$ is a topological invariant defined by the integration of the Berry curvature over the entire Brillouin zone, characterizing the band topology [5]. The value of $C$ determines the number of chiral edge states, while its sign determines the propagation direction of these edge states, or the chirality. In the conventional integer quantum Hall effect, electronic states form Landau levels under high magnetic fields, allowing $C$ to be tuned consecutively from negative to positive values through changes in external magnetic fields or carrier density [6]. However, tuning $C$ in a Chern insulator is generally challenging because the band topology is remarkably robust against external perturbations. Nevertheless, the ability to *in-situ* tune $C$, especially to reverse its sign, is crucial for realizing complex topological device architectures and holds promising applications in topological circuits [7-10]. To date, Chern insulators have been observed in several systems, including magnetically doped topological insulator thin film [2], $MnBi_2Te_4$ [11], twisted graphene [3,12], $WSe_2/MoTe_2$ moiré superlattices [13], twisted $MoTe_2$ [14-17], and rhombohedral graphene [18-24]. However, *in-situ* switching the sign of nonzero $C$ in these Chern insulators remains challenging.

Rhombohedral multilayer graphene has recently emerged as a promising platform for exploring correlated Chern insulators [18-23]. The low-energy bands in rhombohedral multilayer graphene exhibit a unique power-law energy dispersion relation governed by $E \sim k^N$, where $N$ is the layer number and $k$ is the crystal momentum. This dispersion implies the presence of a rather flat band at low energy, with the flatness increasing with $N$, thus fostering strong correlations [20-22,25-34]. Furthermore, the flat surface band in rhombohedral multilayer graphene possesses nontrivial topology with valley-contrasting Chern numbers that are tunable by an electric field, providing the essential ingredients for engineering correlated Chern states [35]. When a moiré superlattice is introduced by aligning rhombohedral multilayer graphene to hexagonal boron nitride (h-BN), it hosts an isolated topological flat band with a nonzero Chern number. Consequently, Chern insulators with $C = 1$ and $C = 2$ have been reported in rhombohedral pentalayer [18] and trilayer graphene [19], respectively. However, switching the sign of $C$ has not yet been achieved. Theoretical calculations suggest that the Chern states in rhombohedral graphene moiré superlattices can be effectively tuned by an electric field, with switchable $C$ between multiple nonzero values [36].

In this work, we report the observations of Chern insulators in singly aligned rhombohedral multilayer graphene/h-BN moiré superlattices. By applying a large displacement field ($D$) in heptalayer (7L) graphene, we identify a cascade of topological phase transitions between multiple nontrivial states ($C$ =-1, 1, 2) at the quarter electron filling of the moiré band, obeying the Streda formula. The Chern states are further stabilized by integrating monolayer tungsten diselenide (WSe2), which provides strong spin-orbit coupling (SOC), into the moiréless interface of rhombohedral decalayer (10L) graphene superlattices. This integration enables the observation of quantized anomalous Hall resistance with $R_{xy} \sim h/2e^2$ down to zero magnetic field, accompanied by pronounced magnetic hysteresis. Displacement-field-driven topological phase transition from $C = -1$ to $C = 2$ has been observed in rhombohedral 10L graphene superlattices. Notably, the ability to switch the sign of the Chern number ($C = -1$ to $C = 1$ in 7L or $C = 2$ in 10L) demonstrates that the chirality of the Chern insulators can be effectively controlled by $D$. Our findings reveal highly tunable topological states in rhombohedral multilayer graphene moiré superlattices.

## II. RESULTS
### A. Phase diagram of rhombohedral 7L graphene moiré superlattices

Our Device D1 was fabricated using h-BN encapsulated rhombohedral 7L graphene, as shown in Fig. 1(a). We employed dual-gate structures to independently tune the carrier density ($n$) and $D$. During the fabrication, the top h-BN was crystallographically aligned with graphene (see Supplemental Materials Fig. S1). Figure 1(d) shows the $n - D$ map of the longitudinal resistance $R_{xx}$ at the base temperature $T = 50$ mK and $B = 0$ T. We observed multiple correlated insulators on the electron-doping side, exhibiting a remarkable asymmetry with respect to $D$. For $D > 0$, a cascade of insulating states appears at moiré filling factors of $\nu = \frac{n}{n_0} = 0, 1, 2, 3, 4$ ($n_0$ is the density corresponding to one electron per moiré unit cell). By comparison, for $D < 0$, strong insulating states are observed only at $\nu = 0$, with weak resistance peaks at $\nu = 1, 2$. This asymmetrical feature suggests a single alignment between graphene and h-BN. The twist angle of the graphene/h-BN moiré superlattice is estimated to be 0.86°, derived from the carrier density at full filling ($\nu = 4$ at $D > 0$). Notably, compared to non-aligned devices, the layer antiferromagnetic (LAF) insulating states occurring near $n = 0$ cm$^{-2}$ and $D = 0$ V nm$^{-1}$ are absent in our singly aligned device (see Supplemental Materials), indicating that the moiré potential weakens the coupling between the top and bottom layer [22,28,30]. In contrast, LAF states are preserved in singly aligned pentalayer graphene devices [18], suggesting that layer number significantly influences the electronic states in rhombohedral graphene superlattices. A similar absence of LAF was observed in doubly aligned rhombohedral 7L graphene (see Supplemental Materials Fig. S2).

Although the moiré superlattice and large layer number in our devices seem to decouple the two surface layers, the moiré potential at the top interface can still influence the electronic states at the bottom interface even with large $|D|$. We found at the region between $D = -0.750$ V nm$^{-1}$ and $D = -0.950$ V nm$^{-1}$, where the electrons are polarized aways from the moiré interface, the weak moiré potential acting on the distant surface can induce multiple topological phase transitions. In contrast, if the moiré potential is strong, either occurring on the $D > 0$ side (see Supplemental Material Fig. S4) or in a doubly aligned device, only topological trivial phases were observed [30].

Figures 1(e) and 1(f) show enlarged $n - D$ maps of symmetrized $R_{xx}$ and anti-symmetrized Hall resistance $R_{xy}$ at $-1.0 < D < -0.7$ V nm$^{-1}$, measured at $B = \pm 0.5$ T. At $\nu = 1$, we observed unusual large $R_{xy}$ values, accompanied by local minima in $R_{xx}$. Additionally, the sign of $R_{xy}$ reverses with increasing $|D|$, while maintaining their large absolute values. In the following sections, we focus on this region and demonstrate that these transitions arise from topologically nontrivial transitions involving changes in Chern numbers.

**B. Cascade of topological phase transitions at $\nu = 1$**

To unveil the origin of the anomalous $R_{xy}$ and its tunability by $D$, we performed fan diagrams by measuring $R_{xx}$ and $R_{xy}$ as a function of $\nu$ and $B$ at various fixed $D$. Figure 2 shows the fan diagrams in the vicinity of $\nu = 1$ at four representative values of $D$. Performance at other $D$ can be found in Supplemental Material Fig. S6. The remarkable features in Fig. 2 are that at all four $D$, $R_{xy}$ exhibits large values at low field regions fanning out from $\nu = 1$, accompanied by corresponding dips in $R_{xx}$. Additionally, $R_{xy}$ reverses its signs at $B = 0$. The dispersions of the local maximum $R_{xy}$ and the corresponding dips in $R_{xx}$ as a function of $B$ follow the Streda formula $\frac{\partial n}{\partial B} = C \frac{e}{h}$, yielding various values of $C$ marked in Fig. 2 [37].

The emergence of the extracted nonzero $C$ utilizing the Streda formula are hallmarks of correlated Chern insulators [21,38-44]. We exclude the possibility of quantum Hall states giving rise to the nonzero $C$ for the following reasons. On the one hand, at a given $D$, only specific Chern states emerge at low field regimes. This is significantly different from the quantum Hall states, where usually a series of $C$ simultaneously appear, and high magnetic field is necessary to form the Landau levels. Indeed, we observe a series of quantum Hall states fanning out from $\nu = 0$ and $\nu = 1$, gradually appearing at $B > 4$ T (see Supplemental Material Fig. S8). On the other hand, within the same device, at $D > 0$ side, we observe topological trivial states showing neither the abnormal large $R_{xy}$ nor nonzero $C$ (see Supplemental Material Fig. S4). This suggests the states at $D > 0$ host distinct topology from those at $D < 0$ side.

In Fig. 2, we observe strikingly $D$-dependent Chern states. At $D = -0.760$ V nm[-1], a Chern state with $C = -1$ dominates at low magnetic fields. Another Chern state with $C = 1$ gradually emerges at $B > 2$ T. When decreasing $D$ to -0.840 V nm[-1], the $C = 1$ state extends to low field region and wins in the competition with $C = -1$ state at $B < 0.4$ T. Further decreasing $D$ to -0.900 V nm[-1], a third Chern state with $C = 2$ emerges and participates in the competition. It dominates over the other two states at $B < 1$ T. The $C = -1$ is expelled to high magnetic field region, while $C = 1$ gradually smears. At $D = -0.920$ V nm[-1], the $C = 2$ state completely occupies the entire phase diagram. The coexistence of multiple Chern states with $C = -1, 1, 2$ and their competitions can also be observed by performing $\nu - D$ maps at fixed magnetic fields (see Supplemental Material Fig. S7). The cascade of topological phase transitions at $\nu = 1$ observed in our rhombohedral heptalayer graphene moiré superlattices differs from that in pentalayer graphene, where only $C = 1$ was observed [18]. The electric field tunability of multiple topological nontrivial states, especially the sign of nonzero $C$, offers a unique platform for topological engineering.

**C. Quantized anomalous Hall effects stabilized by WSe₂**

In Device D1, we observe only infantile ferromagnetic states (see Supplemental Material Fig. S14).

To extend the Chern states into zero magnetic fields, we employed a structure of graphene proximitized by a transition metal dichalcogenide layer. Previous studies on twisted bilayer graphene and rhombohedral multilayer graphene have shown that ferromagnetic states and Chern insulators are significantly enhanced when graphene is in contact with a WSe$_2$ or WS$_2$ layer with strong SOC [20,21,45]. In rhombohedral graphene moiré superlattices, Chern states emerge when electrons are polarized away from the moiré interfaces, allowing us to further incorporate the effects of SOC.

To this end, we fabricated a rhombohedral 10L graphene moiré superlattice using a monolayer WSe$_2$ as the top substate at the moiréless interface (Device D2), as shown in Fig. 3(a). At the bottom surface, the graphene was directly in contact and crystallographically aligned with h-BN, forming the moiré superlattices. To unveil the role of SOC induced by WSe$_2$, we also measured the transport behavior of intrinsic rhombohedral 10L graphene moiré superlattices without WSe$_2$ (Device D3). Device D3 and Device D2 share the same stack and underwent the same fabrication process, resulting in the same moiré wavelength with a twist angle of 0.58°, allowing us to directly compare scenarios with and without SOC. We found that intrinsic rhombohedral 10L graphene moiré superlattices (Device D3) did not exhibit any topological features at either the moiré or moiréless interface (see Supplemental Material Fig. S13), unlike rhombohedral 7L graphene. This difference is probably due to the enhanced trigonal warping effect in thicker layers, which can disrupt the correlated flat band characteristics [27].

In contrast, we observe enhanced topological features, manifesting as quantized anomalous Hall effects, in Device D2. Figure 3 presents detailed measurements of $R_{xx}$ and $R_{xy}$ in Device D2 at the base temperature. Similar to the behavior in rhombohedral 7L graphene, topologically nontrivial states emerge when electrons are polarized toward the moiréless interface, where they experience strong SOC from WSe$_2$ through proximity effects. As shown in Fig. 3(b), a local minimum in $R_{xx}$ is observed at $\nu = 1$ in the region 0.690 V nm$^{-1}$< $D$ <0.870 V nm$^{-1}$. In the same region, $R_{xy}$ in Fig. 3(d) shows an anomalous large value down to $|B| = 50$ mT.

To identify the topological states at $\nu = 1$, we fixed $D = 0.814$ V nm$^{-1}$ and scanned the out-of-plane magnetic field $B$ forward and backward. As shown in Fig. 3(e), $R_{xy}$ displays a pronounced hysteresis loop with a coercive field $B_c \approx 20$ mT, indicating spontaneous time-reversal symmetry breaking. The value of $R_{xy}$ is quantized to $\frac{h}{2e^2} \approx 13$ kΩ down to $B \approx 0$ T, consistent with a quantized anomalous Hall state with $C = 2$ at given $D$. This quantization of $R_{xy}$ is further confirmed by sweeping either $\nu$ at a fixed $D = 0.814$ V nm$^{-1}$ or $D$ at a fixed $\nu = 1$, as shown in Fig. 3(h) and Fig. 3(i), respectively. Each instance of quantization is accompanied by a local dip in $R_{xx}$ with a value below 2 kΩ. Temperature-dependent measurements of the hysteresis loops yield a magnetic Curie temperature of 4 K and a thermal activation gap of $\Delta \approx 2.0$ K (see Supplemental Material Fig. S15).

**D. Electric-field driven nontrivial topological phase transition in rhombohedral 10L graphene**
Beyond observing the Chern state with $C = 2$ at $D = 0.814$ V nm$^{-1}$, we also find that the sign of $R_{xy}$ is tunable by $D$, similar to the behavior in 7L graphene (Device D1) shown in Fig. 1(f). As illustrated in Fig. 3(d), $R_{xy}$ undergoes a sign reversal around $D \approx 0.730$ V nm$^{-1}$. While electrical switching of $R_{xy}$'s sign has been previously reported in twisted graphene systems [3,45,46], the underlying mechanisms in our system differ significantly. In twisted graphene systems, the reversible $R_{xy}$ appears as hysteretic behavior when sweeping either $v$ or $D$ at a fixed small $B$, with sign reversals occurring within a fixed $v$ and $D$ range, and the absolute value of the Chern number ($|C|$) remains unchanged. This phenomenon is a dramatic process and arises from gate-induced valley polarization reversal [3], where the orbital magnetization changes sign as different valleys are favored at a given $B$ during each gate sweep. In contrast, our system exhibits distinct characteristics that point to different mechanisms. First, no hysteresis is observed when sweeping either $v$ or $D$, as shown in Fig. 3(h) and 3(i). Second, $R_{xy}$ reverses its sign at distinct $D$ regions, with $R_{xy}$ positive in the range 0.690 V nm$^{-1}$< $D$ <0.730 V nm$^{-1}$ and negative in 0.730 V nm$^{-1}$< $D$ <0.870 V nm$^{-1}$, as illustrated in Fig. 3(d) and 3(i). Third, a series of fan diagrams stemming from $v = 1$ were measured, from which Chern numbers were extracted by the Streda formula.

Figure 4 shows the evolution of topological nontrivial states as a function of $D$. Obviously, the Chern state at $v = 1$ is highly tunable by both $D$ and $B$. Particularly, it undergoes a topological phase transition from $C = -1$ to $C = 2$ induced by tuning $D$ at $B = 0$. These two distinct topological states exhibit pronounced magnetic hysteresis when sweeping $B$ forward and backward at fixed $v$ and $D$, as seen in Fig. 4(e)-4(h), confirming their spontaneous time-reversal symmetry breaking. Notably, both the sign and absolute value of $C$ are tunable by $D$. Combining the data from Fig. 2 and Fig. 4, we observe sign reversals of $C$ in both 7L and 10L graphene, suggesting that electrical switching of chirality in rhombohedral graphene Chern insulators is not uncommon.

## III. DISCUSSION

Our findings reveal a cascade of topological phase transitions simply by tuning $D$, highlighting the rich topological phases in rhombohedral multilayer graphene moiré superlattices. The nature of Chern insulators at $v = 1$ remains elusive, despite many theoretical studies aimed at understanding the underlying mechanisms [18,47-51]. The gapped states at $v = 1$ are believed to arise from electron interactions, as single-particle calculations consistently predict gapless states [48]. Applying $D$ can redistribute the Berry curvature within the first moiré conduction band at the single-particle level [52]. When strong interactions open a gap, the Berry curvature integrated up to the Fermi level at $v = 1$ can yield different integer Chern numbers for different values of $D$, resulting in observable $D$-induced topological phase transitions. Moreover, the specific Chern numbers may vary with the layer number of rhombohedral graphene, as the Berry curvature distributions differ slightly at the single-particle level. Other mechanisms, such as anomalous Hall crystals, could also contribute to the nontrivial topological phases at $v = 1$ [49,50].

Compared to previous reports of electric-field-tunable Chern insulators in various graphene-based systems [3,23,24,46,53], our work demonstrates the ability to switch the sign of Chern numbers via topological phase transitions, offering an essential ingredient for designing topological circuits. Notably, edge current chirality switching in Chern insulators has recently been achieved in magnetically doped topological insulators via spin-orbit torque assisted by an applied magnetic field [8], which is an extrinsic approach. In contrast, the chirality switching in rhombohedral graphene

Chern insulators here is an intrinsic effect, achieved purely by electric-field tuning, making it a more practical approach for topological electronics.

Our observations of quantized anomalous Hall resistance with a high Chern number ($C = 2$) also offer new avenues for exploring emerging physics and low-power-consumption electronic applications [3,20]. Particularly, given that rhombohedral multilayer graphene moiré superlattices can support fractional Chern insulators [18,47], our system may host tunable fractional Chern insulators with high Chern numbers. We have already observed preliminary features of the anomalous Hall effect at $v < 1$ (as shown in Supplementary Materials Fig. S17), though it is not yet quantized.

While we have achieved quantized anomalous Hall effects for the $C = 2$ state, the quantization of the $C = -1$ state in both 7L and 10L graphene remains incomplete at zero magnetic fields. The lack of full quantization in $R_{xy}$ and the residual $R_{xx}$ may be due to several factors. Unlike in pentalayer graphene [18], where only a single Chern insulator state was observed, multiple Chern insulator states exist in our system. At zero magnetic field, these competing topologically nontrivial states ($C = -1, 1, 2$) are energetically close. As shown in Fig. 2 and Fig. 4, the Chern states are highly sensitive to $D$, so even tiny spatial variations in $D$, due to non-uniform gating, can create different Chern states across adjacent domains. This multi-domain structure can lead to bulk dissipation and percolation transport, disrupting perfect Hall quantization [54]. Applying a magnetic field can localize disorder, facilitate time-reversal symmetry breaking, and enlarge the topological gap, thus improving quantization, as observed in magnetic topological insulators and magic-angle twisted bilayer graphene [38-42,55-57]. Other possible reasons for imperfect quantization include disorder from interfacial charges, twist angle inhomogeneities, or metallic grain boundaries between domains with slightly different twist angles. Additionally, reducing the electronic temperature could further enhance quantization.

In summary, our findings demonstrate that rhombohedral multilayer graphene moiré superlattices provide a fertile platform for topological engineering. The tunability of Chern numbers in these Chern insulators offers new opportunities to design topological junctions. With multiple tuning parameters, such as twist angle, strength of SOC, and layer number, rhombohedral graphene moiré superlattices allow for versatile approaches to engineer diverse topological phases. Our observation of Chern number switching enables the creation of Chern junctions and the development of topological circuits, where domains with different Chern numbers can be controlled via multiple local gates [7].

## IV. METHODS
### A. Device fabrication
Multilayer graphene was mechanically exfoliated from bulk natural graphite (Graphenium Flakes, NGS Naturgraphit) onto $SiO_2$ (285 nm)/Si substrates. The bulk $WSe_2$ crystal was grown using chemical vapor transport methods. The layer number of graphene flakes was initially determined by optical contrast and further confirmed by transport measurements (see Fig. S3). Rhombohedral stacking domains were hunted by performing Raman maps of graphene. To stabilize the rhombohedral stacking orders in subsequent van der Waals assembly processes, we isolated them from Bernal stacking domains by a cutting process using a tungsten tip. The h-BN encapsulated rhombohedral graphene stacks were prepared using a standard dry transfer method with the assistance of poly(bisphenol A carbonate)/polydimethylsiloxane (PC/PDMS) stamp. For Device D1, during the transfer, we intentionally aligned one of the straight edges of top h-BN with that of the graphene [see Fig. S1(a)] and misaligned them with that of the bottom h-BN, resulting in a single alignment configuration. For Device D2 and D3, we first picked up monolayer WSe$_2$ by the top h-BN, followed by picking up a rhombohedral graphene flake. The heterostructures were released onto bottom h-BN/Pt bottom gate, which was prepared in advance. The graphene was intentionally aligned with the bottom h-BN. The alive rhombohedral stacking domains in the final stacks were identified by a second Raman map, as demonstrated in Fig. S1(d). We further used atomic force microscopy to select bubble free regions to make devices. For device fabrication, standard e-beam lithography, inductively coupled plasma, and e-beam evaporation were employed to design a multi-terminal Hall bar geometry. Contact electrodes were made by selectively etching the top h-BN and evaporating Cr/Au (3 nm/60 nm) on top of the exposed multilayer graphene.

**B. Transport measurement**

Low-temperature electronic transport measurements were conducted in a dilute refrigerator (Oxford Triton) with a base temperature of 50 mK. Standard low-frequency lock-in technique (SRS 830) at a frequency of 17.7 Hz was used to measure the longitudinal and Hall resistance of a Hall bar device. The AC excitation current was set to be 1-10 nA. Gate voltages were applied using a Keithley 2614B.

Dual-gate structures were utilized to independently tune the total carrier density $n$ and the displacement field $D$. Their values were converted from the top gate ($V_t$) and bottom gate ($V_b$) voltages using a parallel plate capacitor model: $n = \frac{C_b \Delta V_b}{e} + \frac{C_t \Delta V_t}{e}$ and $D = \frac{C_b \Delta V_b - C_t \Delta V_t}{2\varepsilon_0}$, where $C_b$ ($C_t$) are the bottom (top)-gate capacitances per unit area, $\Delta V_b = V_b - V_b^0$ ($\Delta V_t = V_t - V_t^0$) are the effective bottom (top) gate voltages, $e$ is the elementary charge, and $\varepsilon_0$ is the vacuum permittivity. $C_b$ ($C_t$) were measured from the Hall effect at normal regions and calibrated by Landau level sequences at $\nu = 0$ (see Fig. S8).

**C. Symmetrized $R_{xx}$ and anti-symmetrized $R_{xy}$**

The measured Hall resistance $R_{xy}$ inevitably contains signals from longitudinal resistance $R_{xx}$ due to the imperfect Hall bar geometry. To eliminate the components of $R_{xx}$ from $R_{xy}$, we used the standard procedure to obtain symmetrized $R_{xy}$ and anti-symmetrized $R_{xy}$. For the data shown at a fixed $B$ [such as Figs. 1(e) and 1(f)], we used: $R_{xy}(\pm B) = [R_{xy}(B) - R_{xy}(-B)]/2$ and $R_{xx}(\pm B) = [R_{xx}(B) + R_{xx}(-B)]/2$. For magnetic hysteresis data [such as Fig. 3(e)], we employed: $R_{xy}^{\text{antisym}}(B, \leftarrow) = [R_{xy}(B, \leftarrow) - R_{xy}(-B, \rightarrow)]/2$ and $R_{xy}^{\text{antisym}}(B, \rightarrow) = [R_{xy}(B, \rightarrow) - R_{xy}(-B, \leftarrow)]/2$, where $\leftarrow$ and $\rightarrow$ indicate the magnetic field sweep direction.

**ACKNOWLEDEGMENTS**

This work was funded by National Natural Science Foundation of China (Grant No. 12274354, S.X.), the Zhejiang Provincial Natural Science Foundation of China (Grant No. LR24A040003, S.X.; XHD23A2001, S.X.), the R&D Program of Zhejiang province (Grant No. 2022SDXHDX0005,

W.Z.), and Westlake Education Foundation at Westlake University. We thank Chao Zhang from the Instrumentation and Service Center for Physical Sciences (ISCPS) at Westlake University for technical support in data acquisition. We also thank Westlake Center for Micro/Nano Fabrication and the Instrumentation and Service Centers for Molecular Science for facility support. K.W. and T.T. acknowledge support from the JSPS KAKENHI (Grant Numbers 21H05233 and 23H02052) and World Premier International Research Center Initiative (WPI), MEXT, Japan. Z.S. was supported by project LUAUS23049 from Ministry of Education Youth and Sports (MEYS) and by the project Advanced Functional Nanorobots (reg. No. CZ.02.1.01/0.0/0.0/15_003/0000444 financed by the EFRR).

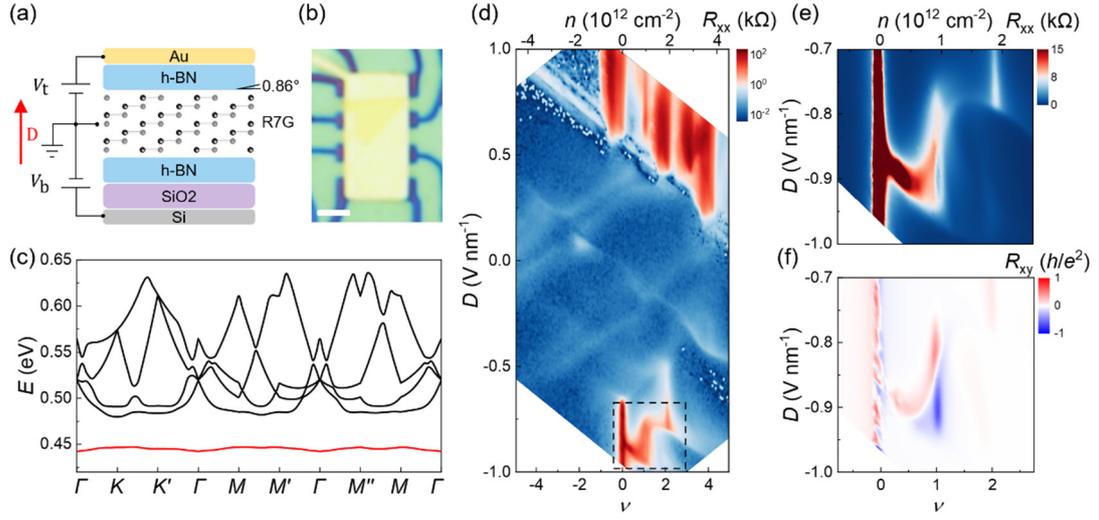

FIG. 1. Phase diagram of rhombohedral heptalayer graphene moiré superlattice (Device D1). (a) Schematic of the device structure. The moiré superlattice is formed at the interface between graphene and top h-BN. The red arrow illustrates the direction of the positive $D$. (b) Optical image of the final device with a standard Hall bar geometry. The scale bar is 2 μm. (c) Topological flat band of rhombohedral heptalayer graphene moiré superlattices calculated based on Hartree-Fock method. The calculated Chern number of the isolated band (red color) has a nonzero number of 1, indicating the topological nontrivial characteristic. (d) Longitudinal resistance $R_{xx}$ as a function of the filling factor $\nu$ and $D$ at $B = 0$ T. The top $x$-axis labels the corresponding carrier density $n$. The black dashed box highlights the region of interest in the main text. (e) (f) Zoomed-in phase diagram in the region highlighted by the black dashed box in (d), showing the symmetrized $R_{xx}$ (e) and anti-symmetrized Hall resistance $R_{xy}$ (f) as a function of $\nu$ and $D$. The data were measured at $B = \pm 0.5$ T. At $\nu = 1$, anomalously large $R_{xy}$ emerge with both positive and negative signs tunable by $D$. All the data were measured at the base temperature of $T = 50$ mK.

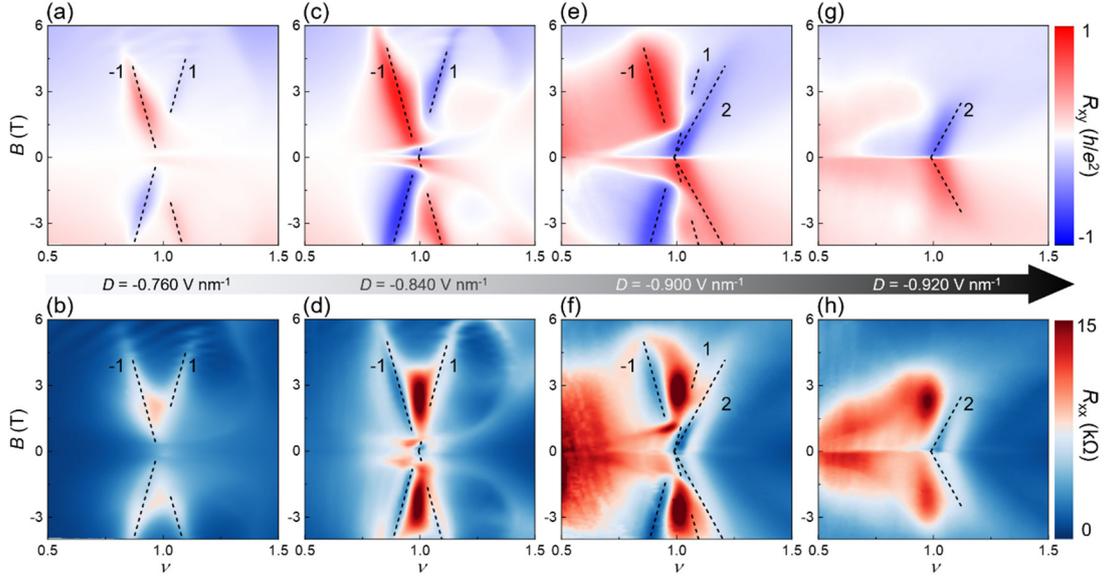

FIG. 2. Electric- and magnetic-field driven a cascade of topological phase transitions in Device D1. (a) (b) $R_{xy}$ (a) and $R_{xx}$ (b) as a function of $v$ and $B$ at a fixed $D = -0.760$ V nm$^{-1}$. The dashed lines mark the Chern numbers of the topological nontrivial states $C = -1$ and $C = 1$. The $C = -1$ state dominates over $C = 1$ at low $B$ region. (c) (d) $R_{xy}$ (c) and $R_{xx}$ (d) as a function of $v$ and $B$ at a fixed $D = -0.840$ V nm$^{-1}$. The $C = 1$ state becomes dominant at low $B$ region. (e) (f) $R_{xy}$ (e) and $R_{xx}$ (f) as a function of $v$ and $B$ at $D = -0.900$ V nm$^{-1}$. A third Chern state with $C = 2$ emerges from low field regions and participates the competition with the $C = 1$ and $C = -1$ states. (g) (h) $R_{xy}$ (g) and $R_{xx}$ (h) as a function of $v$ and $B$ at $D = -0.920$ V nm$^{-1}$. The Chern state with $C = 2$ wins the competition and dominates in the entire phase diagram. All Chen numbers are determined according to the Streda formula.

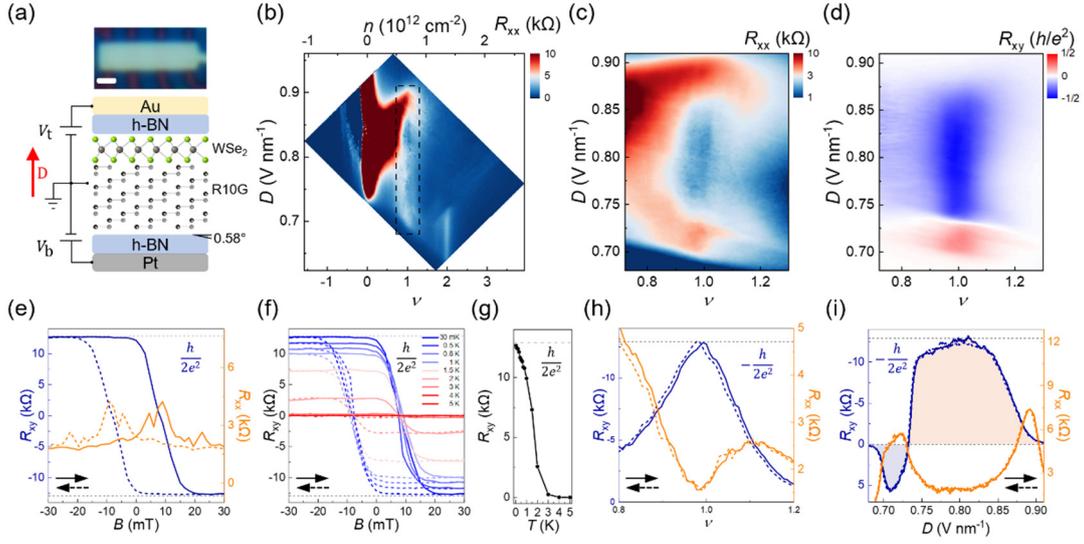

FIG. 3. Phase diagram and quantized anomalous Hall effects in rhombohedral 10L graphene moiré superlattice stabilized by WSe$_2$ (Device D2). (a) Optical image (top panel) and schematic (bottom panel) of the device. The scale bar is 1 μm. The moiré superlattice is formed at the interface between graphene and bottom h-BN. (b) $R_{xx}$ as a function of $\nu$ and $D$ at $B=0$ T in the region where electrons are polarized to the moiréless interface. The black dashed box highlights the region of interest discussed in the main text. (c) (d) Symmetrized $R_{xx}$ (c) and Anti-symmetrized $R_{xy}$ (d) as a function of $\nu$ and $D$ at $B=\pm 50$ mT. The data were acquired in the region marked by the black dashed box in (b). (e) $R_{xy}$ and $R_{xx}$ as a function of $B$ swept forward and backward across zero at fixed $\nu=1$ and $D=0.814$ V nm$^{-1}$. (f) Temperature dependent magnetic hysteresis loops acquired by sweeping $B$ forward and backward at various temperatures. (g) Extracted $R_{xy}$ from the data in (f) as a function of temperature at $B=\pm 30$ mT. (h) $R_{xy}$ and $R_{xx}$ as a function of $\nu$ swept forward and backward at a fixed $D=0.814$ V nm$^{-1}$. (i) $R_{xy}$ and $R_{xx}$ as a function of $D$ swept forward and backward at a fixed $\nu=1$. The different colored regions correspond to opposite signs of the Chern numbers. (h) and (i) were measured at a fixed $B=\pm 50$ mT.

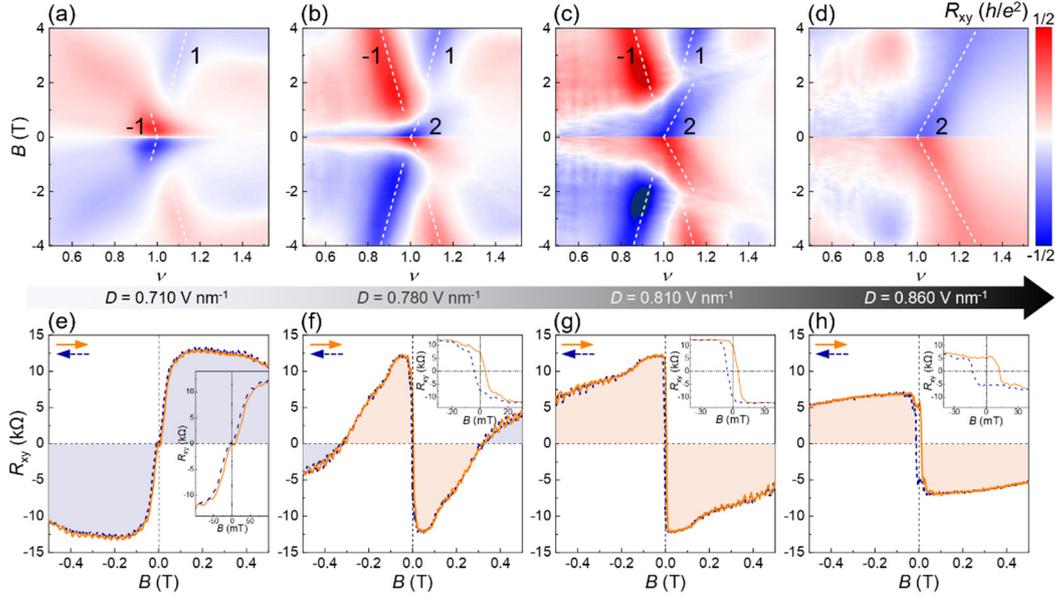

FIG. 4. Nontrivial topological phase transitions in rhombohedral 10L graphene proximitized by WSe$_2$ (Device D2). (a)-(d) $R_{xy}$ as a function of $\nu$ and $B$ at various fixed $D$. The dashed lines mark the Chern numbers of topological nontrivial states following the Streda formula. (e)-(h) $R_{xy}$ as a function of $B$ swept forward and backward across zero at fixed $\nu = 1$ and corresponding $D$. The colored regions correspond to opposite signs of the Chern numbers, with light red (blue) corresponding to positive (negative) values. The insets in each panels show enlarged views of the hysteresis loops in low $B$ regions.